\documentclass[floatfix,showpacs]{revtex4}
\usepackage{times}
\usepackage{epsfig}
\usepackage{amssymb}
\usepackage{amsmath}
\usepackage[english]{babel}
\usepackage{graphics}

\begin{document}

\bibliographystyle{apsrev}

\title{Structural Evolution of the Brazilian Airport Network}

\author{Luis E. C. da Rocha}
\email{luis.rocha@tp.umu.se}

\affiliation{Department of Theoretical Physics \\
Ume{\aa} University, Ume{\aa}, Sweden}

\date{24th March 2009}

\begin{abstract}

The aviation sector is profitable, but sensitive to economic fluctuations, geopolitical constraints and governmental regulations. As for other means of transportation, the relation between origin and destination results in a complex map of routes, which can be complemented by information associated to the routes themselves, for instance, frequency, traffic load or distance. The theory of networks provides a natural framework to investigate dynamics on the resulting structure. Here, we investigate the structure and evolution of the Brazilian Airport Network (BAN) for several quantities: routes, connections, passengers and cargo. Some structural features are in accordance with previous results of other airport networks. The analysis of the evolution of the BAN shows that its structure is dynamic, with changes in the relative relevance of some airports and routes. The results indicate that the connections converge to specific routes. The network shrinks at the route level but grows in number of passengers and amount of cargo, which more than doubled during the period studied.

\end{abstract}

\pacs{89.40.Dd,89.75.Hc,89.75.Fb}
\maketitle

\section{Introduction}

During the $20th$ century, society witnessed a large change in terms of transportation options. In the first half of the century, transportation was dominated by ships in intercontinental routes while trains were the most important carriers in continental routes. The aviation sector developed at a fast pace during these years, nowadays it dominates intercontinental traffic and it has a strategic importance in continental flows. This sector has an increasing participation in the domestic transportation market, competing with other more traditional transportation categories. The short travel time is the major motivation to choose air transportation, even though the flight prices are still high in some regions. The flight velocity together with long range routes have deep implications in terms of transportation. People can move fast between different regions and consequently, spread information and diseases faster as well. It was already observed that the world wide airport network is of importance for global epidemics~\cite{Colizza:06}.

The development of the air transportation naturally results in a complex map of routes established between airports. These routes can be regular or non-regular and they are mainly constrained by economic, geopolitical, regulatory and technological factors. This complex structure of flights can be mapped into a network and analyzed by using tools from the so-called complex network theory~\cite{Boccaletti:review, Costa:review,Costa2:review}. Indeed, the airports network structure has been investigated before on the world wide~\cite{Colizza:06,Barrat:04,Barthelemy:05,Guimera:05,Costa2:review} and domestic levels~\cite{Li:04,LiPing:03,Bagler:04,Chi:04,Li:06,Wang:05,Han:07,Costa2:review}, involving countries with different economic and politic situations, and population/area sizes. Some attempts to model the world airport network (WAN) have been proposed considering geopolitical constraints~\cite{Guimera:04}, passengers behavior~\cite{Hu:07} and optimization principles~\cite{Barthelemy:06}.

In particular, the WAN has skewed distributions for degree, passengers traffic and betweenness centrality in an extended range~\cite{Barrat:04,Barthelemy:05,Guimera:05}. In a weighted version, Barrat and collaborators found a strong correlation between the number of routes and the passenger traffic in an airport, and a linear relation between the average passengers traffic and the betweenness~\cite{Barrat:04,Barthelemy:05}. Using another database, Guimer\`a et al. suggested that there are some cities with a small number of routes but large betweenness which could be related to geopolitical constraints~\cite{Guimera:05}. Moreover, they found a hierarchical organization where less connected nodes generally belong to well interconnected communities, while hubs connect many airports that are not directly connected~\cite{Guimera:05}. In the case of the Chinese and US airport networks, different attempts were made to fit the degree distribution which in principle, obeys double Pareto's law~\cite{Li:04,LiPing:03,Li:06}. On the other side, the Indian and Austrian networks apparently follow a single power-law~\cite{Bagler:04,Han:07}.

Although time evolution analysis has been already performed in a short time-scale basis~\cite{Li:04,LiPing:03,Wang:05,Han:07} and shows that there are only small fluctuations in some structural features of the airport network during the week, the time evolution analysis in a year-scale has not been performed yet. The importance of such study is related to the long term evolution of the airport network structure. Such structural evolution reflects the economic health of the system (air transportation related companies and the country itself), since routes are inserted and removed according to passengers and cargo demand, and new airports are built and expanded according to regional development. The network evolution is also constrained by governmental policies which regulate the sector~\cite{Febeliano:06} and by economic fluctuations (e.g. oil prices) which directly affect transportation sectors such as aviation~\cite{Febeliano:06}.

The following analysis of the evolution of the Brazilian airport network uncovered a number of features, suggesting that despite of small fluctuations in some structural properties, the network has undergone a strong rewiring dynamics in the considered period. What in principle might appear as a slowdown of the aviation sector, in fact is a reorganization of the structure according to changing demand. One of the consequences has been an increase in the centrality of some airports, which might affect the performance of the network.

\section{The Brazilian Airport Network}

Extensive literature on the fundamentals of complex networks is available~\cite{Boccaletti:review, Costa:review,Costa2:review}, however, some basic definitions will be shortly introduced before the discussion of the results. A network $\Gamma:=(V,E)$ can be defined as a set $V$ of nodes $i$ ($i=1,2,...,N$) and a set $E$ of $K$ edges $(i,j)$ connecting these nodes according to some rule. Another set of real numbers $W$ is also defined, such that one or more elements of this set can be mapped into each edge as weights, in order to represent a specific feature of the system. Representing the network as a matrix $A$ (adjacency matrix), the element $a_{ij}$ is $1$ if there is an edge connecting nodes $i$ and $j$ (the direction of the edges can be distinguishable). Another matrix $W$ (weight matrix) can be defined such that each element carries information about the weights assigned to the respective edges. By these definitions, the degree (total number of edges or direct neighbors) and the strength (total weight of the edges) of a node $i$ are respectively $k(i)=\sum_{j}a_{ij}$ and $s(i)=\sum_{j}a_{ij}w_{ij}$. In a directed network, not all edges are bidirectional, which means that $(i,j)$ is not necessarily the same as $(j,i)$. To quantify this asymmetry, we define the reciprocity $R$ to count the proportion of bidirectional edges in the network (eqn.~\ref{eq:01}).

\begin{equation}
\label{eq:01}
  R \equiv \dfrac{\sum_{i\neq j}^N (a_{ij}-\overline{a})(a_{ji}-\overline{a})}{\sum_{i\neq j}^N (a_{ij}-\overline{a})^2} \\
\end{equation}
\begin{displaymath}
  \text{where:}\; \overline{a} \equiv \dfrac{\sum_{i\neq j}^N a_{ij}}{N(N-1)}
\end{displaymath}

A path of size $d$ connecting nodes $i$ and $j$ is defined as a sequence of $d-1$ nodes from the set $V$ such that there is an edge connecting the node $m$ with the node $m+1$ of the sequence, being the first and last nodes in the sequence, respectively, $i$ and $j$. If there is a path between $i$ and $j$, it means that $j$ is reachable from $i$. A geodesic of size d ($\sigma_{ij}^d$) is the shortest path between nodes $i$ and $j$. The diameter $D_{dir}$ is defined as the largest geodesic in the network, which is the minimum distance necessary to reach any pair of nodes. The average distance $d_{dir}$ measures the average minimum distance between any reachable pair of nodes (eqn.~\ref{eq:02}).

\begin{equation}
\label{eq:02}
  \langle d_{dir} \rangle \equiv \dfrac{1}{N(N-1)}\sum_{i\neq j} \sigma_{ij}^d
\end{equation}

We can measure the amount of geodesics going through a specific node by using the Freeman betweenness centrality~\cite{Freeman:79}. Given that $\mid \sigma_{ij}(h)\mid$ is the number of geodesics between nodes $i$ and $j$ that pass through node $h$ and $\mid \sigma_{ij}\mid$ is the total number of geodesics between $i$ and $j$, we get the proportion of shortest paths going through specific nodes by using equation~\ref{eq:03}.

\begin{equation}
\label{eq:03}
  B(h) \equiv \dfrac{1}{(N-1)(N-2)}\sum_{i\neq j,i\neq h,j\neq h} \dfrac{\mid\sigma_{ij}^d(h)\mid}{\mid\sigma_{ij}^d\mid}
\end{equation}

\begin{figure}[ht]
  \begin{center}
  \includegraphics[scale=0.21, angle=0]{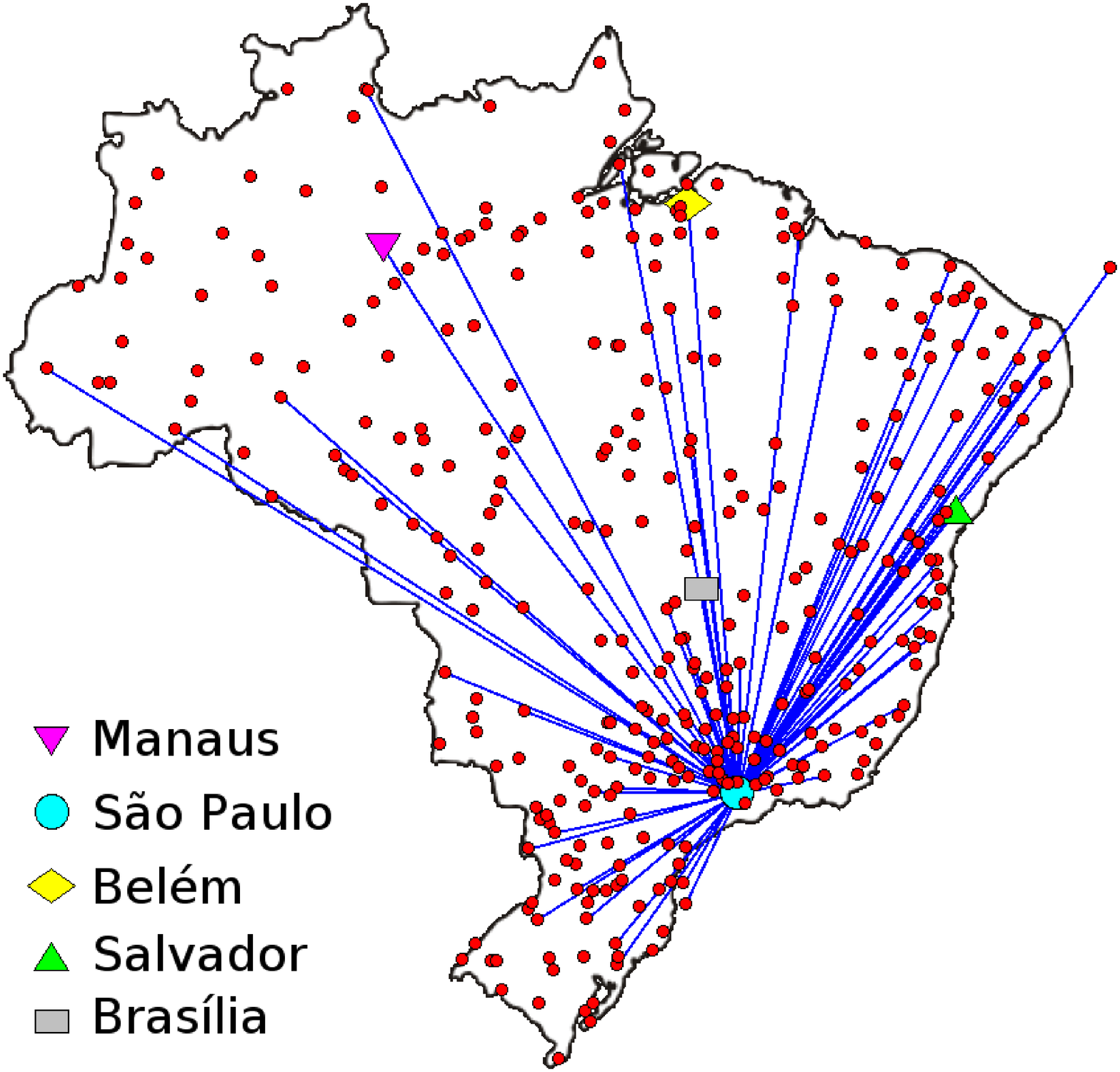}
  \end{center}
  \caption{(Color online) Routes starting/ending in S\~ao Paulo city in $2006$. Some important cities are highlighted, e.g. the capital \textit{Bras\'ilia}. Support material includes the visual evolution of these routes from/to S\~ao Paulo along the years.}  \label{fig:00}
\end{figure}

The \textit{Brazilian airport network} (BAN) is so defined by mapping Brazilian cities (henceforth referred only as airports unless specified) into nodes $i$ and associating a directed edge $(i,j)$ between two nodes $i$ and $j$, whether the respective cities are served by airports containing non-stop regular flights (henceforth named \textit{route} $(i,j)$) (see fig.~\ref{fig:00}). By this definition all the non-regular flights (e.g. air charters, private, military) as well as airports without regular flights are discarded. For each year, from $1995$ to $2006$, a multi-layered network $BAN_{year}$ with $N_{year}$ nodes and $K_{year}$ edges is constructed. Each layer has the same set of nodes and edges, but represents different features associated to the respective route and represents different weights~\cite{Erez05a}. In the present case, the network is defined with $3$ layers: the first ($L^1$) is associated to the number of connections (or flights) completed during the year in the respective route. The second ($L^2$) corresponds to the number of passengers and the third ($L^3$) to the amount of cargo (in kilograms) carried by these flights. In this specific system, the multi-layered approach is equivalent to assigning three different weights for each edge in the network.

The data used to build the BAN are publicly available in the web-page of the \textit{Ag\^encia Brasileira de Avia\c{c}\~ao Civil} (Brazilian National Agency of Civil Aviation) and they are directly extracted from the electronic files of the \textit{Anu\'ario Estat\'istico do Transporte A\'ereo} (Air Transportation Annual Report) which contains specific information about flights and statistics about the domestic and non-domestic airline companies operating in Brazil~\cite{anuar}. The airports are identified either by the name of the main served city ($\> 75\%$) or by the airport name ($\sim 20\%$). Approximately $5\%$ of the airports serve the same metropolitan area, i.e. are settled within the same city or nearby city. Airports within the same metropolitan area are considered as one node in the present network. The data are filtered to remove self-loops, which in principle are meaningless and appear specially in old files for unknown reasons (possibly misprints). Multiple entries are also removed and the respective data are summed into a single entry (except in the cases where misprints are clearly identified). It is important to note that flight information of regional and national companies are separated till $1999$, these data are also summed in order to represent the complete national flight information. All the networks are observed to be completely connected with the exception of one year ($1999$), which besides the giant component, contains $2$ small connected components with $2$ nodes each. This feature has negligible effects in the following analysis and we can safely assume that all networks are connected.

For comparative analysis, the network is randomized maintaining the degree distribution~\cite{Guimera:05} with self-loops and multiple edges being forbidden. Observe that bidirectional edges are not considered as multiple edges. The measurements are obtained by averaging over $10$ network samples. This randomization mechanism is adopted to skip the influence of the degree distribution in the results.

\section{Structure and Evolution}

\subsection{Unweighted BAN}

The configuration of the BAN is observed to be dynamic. Despite some fluctuations, the total number of cities $N_{year}$ (i.e. nodes) and average number of routes $k_{in+out}$ (i.e. directed edges) decreased consistently as observed in Table~\ref{tab:01}. For the sake of comparison with previous results of other airport networks, the average degree $\langle k_{und} \rangle$ and clustering coefficient $\langle cc_{und} \rangle$~\cite{Costa2:review} of the undirected version of the BAN are also presented in Table~\ref{tab:01}. The clustering coefficient measures the probability of two common neighbors of a node $i$ being connected with each other. While the average degree decreased by approximately $28.3\%$ in $2006$ if compared with $1995$, the values of the clustering coefficient are relatively high and stable during the period. In comparison to other airport networks, the BAN is positioned as the second largest national network after the USA when ranked by the number of airports~\cite{LiPing:03,Wang:05}. Moreover, the high clustering coefficient and the small average shortest path length $\langle d_{dir}\rangle$ (eqn.~\ref{eq:01}) indicate that the BAN is small-world, which is in accordance with the other national networks.

In previous papers, the authors usually suggest that with good approximation the direction of the routes are irrelevant~\cite{Li:04,Bagler:04,Barrat:04,Barthelemy:05,Guimera:05} for those networks. Nevertheless, by measuring the reciprocity $R$~(eqns.~\ref{eq:01}) of the BAN, it is clear that although highly reciprocal, $R$ is not close to $1$ (Table~\ref{tab:01}), which means that one can fly from one airport to another but not necessarily return using the same route. Therefore, to maintain this path feature of the network, hereafter the directed version of the network will be adopted.

The average shortest path length $\langle d_{dir} \rangle$~(eqn.~\ref{eq:02}) is in general, slightly higher ($\sim 5\%$) than the randomized version of the same network, while the diameter $D_{dir}$ is about $12\%$ larger. These results might be related to geographical constraints which are unavoidable in the BAN due to the size and the shape of the country, some small (regional) airports are only connected to closer airports and have no long-range routes due to restricted traffic demand. The same effect could explains why the diameter fluctuated little during the period (Table~\ref{tab:01}).

\begin{table*}
\begin{center}
  \begin{tabular}{ | c || c | c | c | c | c | c | c | c | c | c | c | c | }
  \hline
  Year           & 1995 & 1996 & 1997 & 1998 & 1999 & 2000 & 2001 & 2002 & 2003 & 2004 & 2005 & 2006 \\ \hline
  $\centering N$ & $211$  & $224$  & $234$  & $219$  & $208$  & $183$  & $171$  & $172$  & $143$  & $152$  & $154$  & $142$ \\ \hline
  $\centering \langle k_{und} \rangle$ & $13.19$ & $11.49$ & $10.63$ & $10.49$ & $10.46$ & $10.60$ & $11.09$ & $9.79$ & $9.71$ & $9.49$ & $9.73$ & 10.28 \\ \hline
  $\centering \langle cc_{und}\rangle$ & $0.66$ & $0.65$ & $0.62$ & $0.63$ & $0.66$ & $0.66$ & $0.62$ & $0.63$ & $0.62$ & $0.61$ & $0.65$ & $0.63$ \\ \hline
  $\centering \langle k_{in+out}\rangle$ & $22.84$ & $20.37$ & $18.81$ & $18.80$ & $18.86$ & $19.04$ & $19.64$ & $17.70$ & $17.55$ & $17.16$ & $17.65$ & $18.32$ \\ \hline
  $\centering R$ & $0.836$ & $0.865$ & $0.864$ & $0.879$ & $0.885$ & $0.880$ & $0.863$ & $0.888$ & $0.887$ & $0.888$ & $0.891$ & $0.869$ \\ \hline
  $\centering D_{dir}$ & $6$ & $5$ & $6$ & $6$ & $6$ & $5$ & $5$ & $6$ & $6$ & $5$ & $5$ & $5$ \\ \hline
  $\centering \langle d_{dir} \rangle$ & $2.44$ & $2.47$ & $2.36$ & $2.39$ & $2.34$ & $2.49$ & $2.40$ & $2.39$ & $2.36$ & $2.26$ & $2.36$ & $2.34$ \\ \hline
  $\centering S_{in} $ & $3.10$ & $3.09$ & $3.05$ & $3.04$ & $3.01$ & $2.91$ & $3.00$ & $2.93$ & $2.92$ & $2.87$ & $2.91$ & $2.92$ \\ \hline
\end{tabular}
\end{center}
\caption{Evolution of the basic structural features of the BAN network.} \label{tab:01}
\end{table*}

\begin{figure*}[ht]
  \begin{center}
  \includegraphics[scale=0.45, angle=-90]{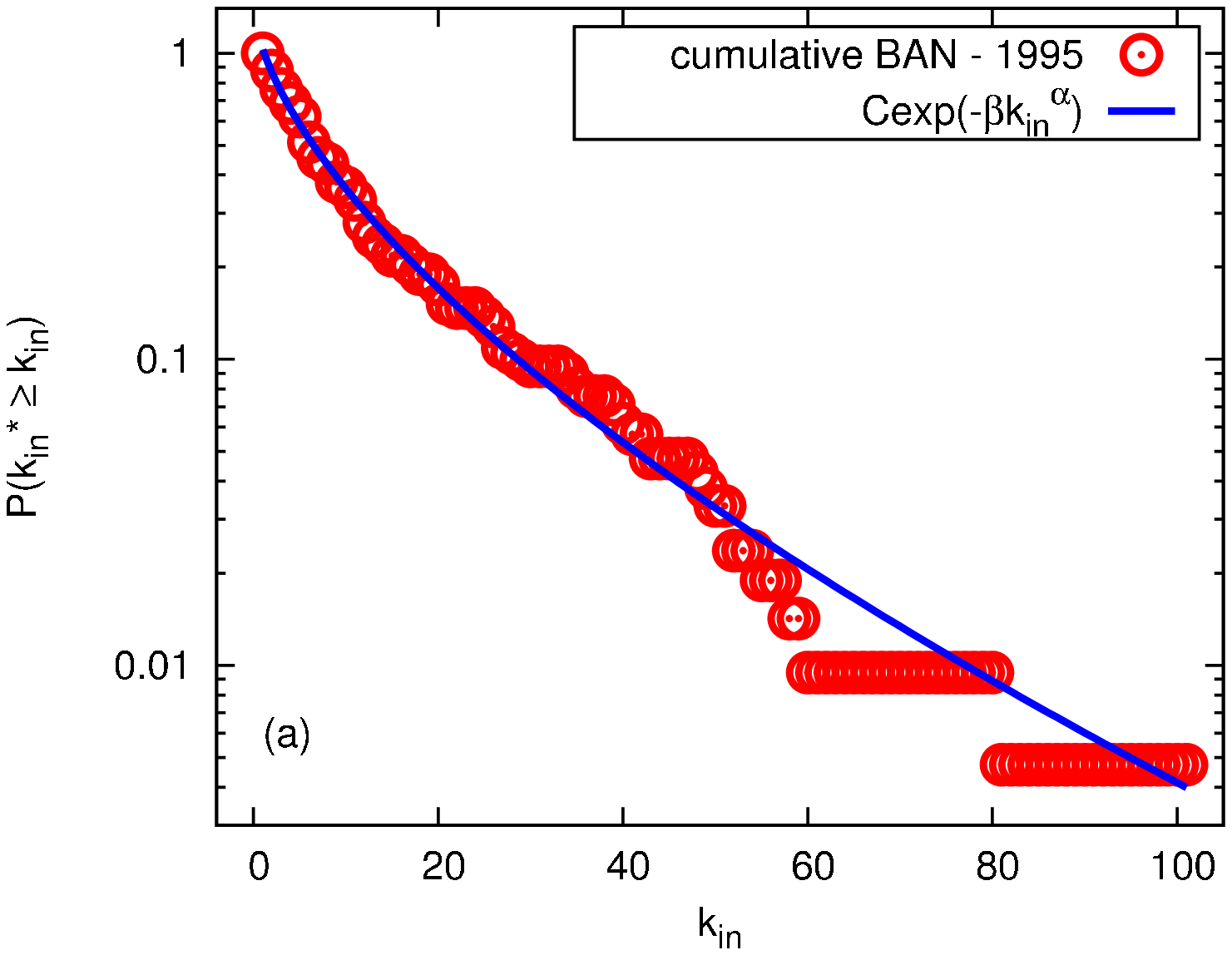}
  \includegraphics[scale=0.45, angle=-90]{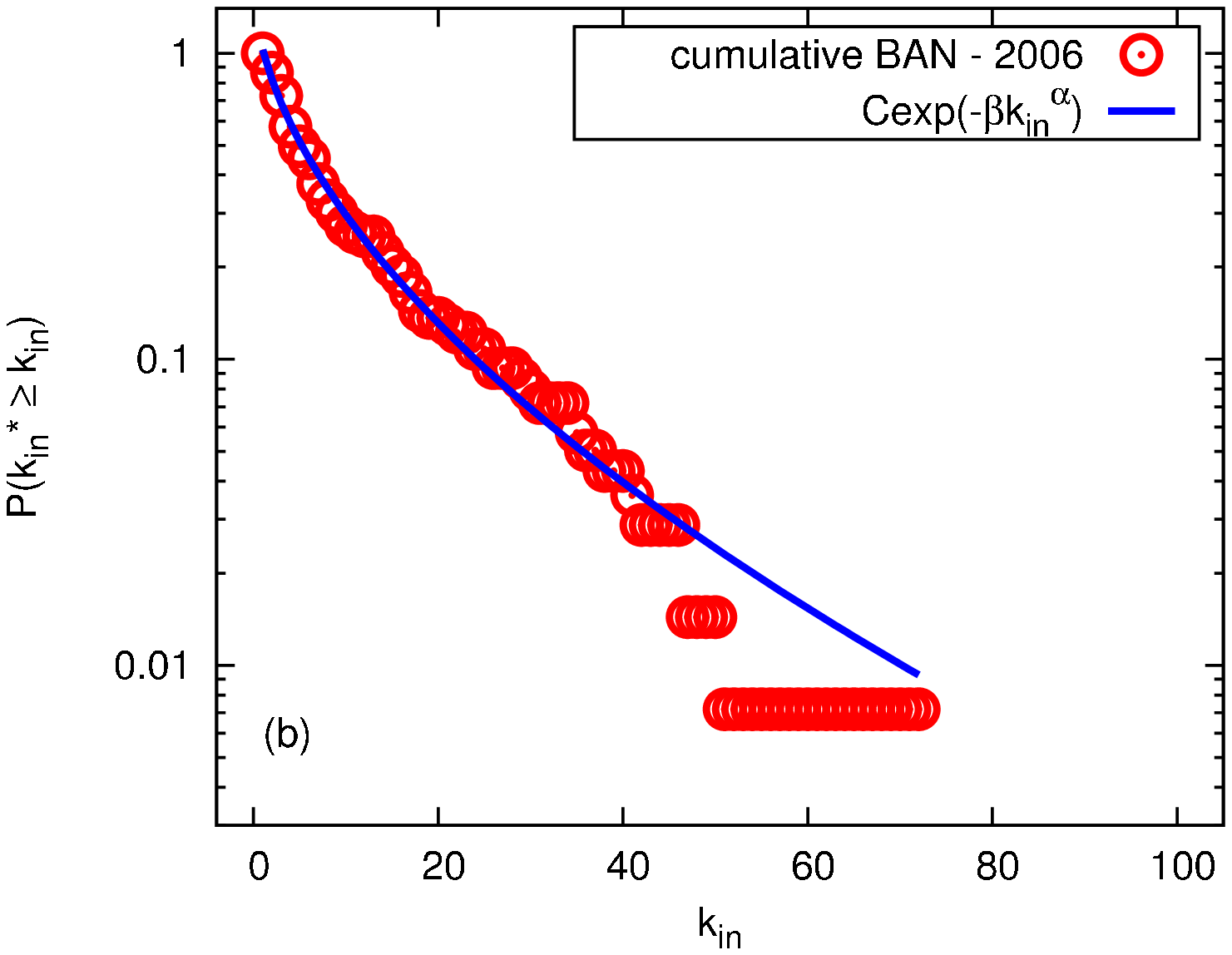} \\
  \end{center}
  \caption{(Color online) Cumulative in-degree distribution of the BAN for (a) $1995$ and (b) $2006$. Reasonable good fits are given by stretched exponential functions with parameters (a) $C=1.414$, $\beta=0.3196$ and $\alpha=0.631$, and (b) $C=1.648$, $\beta=0.474$ and $\alpha=0.559$.}  \label{fig:01}
\end{figure*}

The high correlation (Pearson coefficient $r\sim0.99$) between the number of in- and out-degree permit us to analyze the in-degree distribution and extend the results to the out-degree. Since the BAN is relatively small for statistical purposes, the cumulative data are taken to diminish the data spread, but even so, it is not conclusive which is the right functional form to best describe the in-degree cumulative distribution. A reasonable fit is obtained by using a stretched exponential $P(k*_{in} \geqslant k_{in}) = C\exp(-\beta k_{in}^{-\alpha})$ (Fig.~\ref{fig:01}). We verify that in recent years, some of the most connected airports are operating close or even over the estimated capacity, which means that possibly the decay is a result of physical constraints in the airports as pointed out before~\cite{Barrat:04}.

A better picture of the evolution of the degree heterogeneity (e.g. in-degree) is obtained by measuring the entropy $S_{degree}$ of the distribution $P_{degree}(k)$~\cite{Wang:06} (eqn.~\ref{eq:04}). In this context, small entropy indicates that the network has a well known degree while high entropy show the opposite, i.e. a diversity in the degree values. The results (Table~\ref{tab:01}) suggest that the nodes have became less heterogeneous (lower entropy) with time. As pointed out before, a more homogeneous network is less robust to random failures~\cite{Wang:06}. A \textit{comparative analysis} during the years suggest that the continuous skewed feature of the degree distribution maintain the network vulnerable to hub fails (e.g. congestion), but although small, the increasing homogeneity might contribute to make the network \textit{more} fragile against random failures as well.

\begin{equation}
\label{eq:04}
  S_{degree} \equiv -\sum_{k=1}^{k_{max}} P_{degree}(k)\log_{\epsilon}{P_{degree}(k)}
\end{equation}

\begin{figure}[ht]
  \begin{center}
  \includegraphics[scale=0.45, angle=-90]{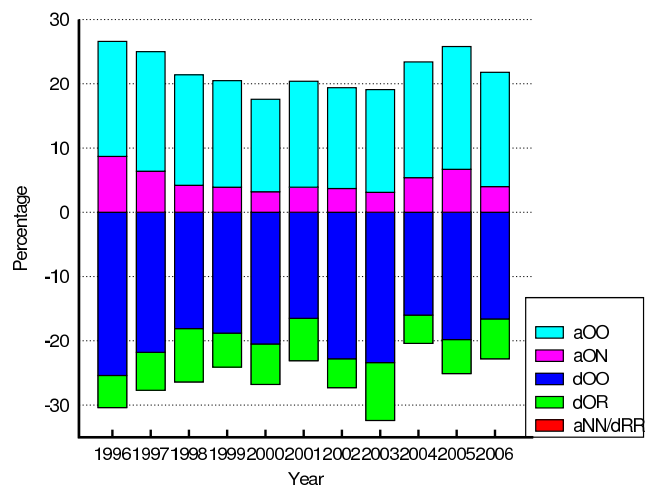}
  \end{center}
  \caption{(Color online) Percentage of added (a) (based in the current year) and deleted (d) (based in the previous year) routes related to new (N), removed (R) and old (O) airports. Case aOO: added routes between old airports; aON: added routes between old and new airports; dOO: deleted routes between old airports; dOR: deleted routes between old and removed airports; aNN: added routes between new airports; and dRR: deleted routes between removed airports.}  \label{fig:03}
\end{figure}

On previous analyses, the authors argue that the world~\cite{Guimera:05} and the Chinese~\cite{Li:04} airport networks have a relatively stationary structure within a period of a year, however, the measurements fluctuations observed so far suggest that a rewiring dynamics is taking place over the years in the BAN. Investigating the establishment of new routes, we observe that a rewiring process underlies the evolution of the BAN, though routes connecting new airports~\footnote{In this paper, \textit{new}, \textit{old} and \textit{removed} airports correspond to \textit{new}, \textit{old} and \textit{removed} nodes in the network. More specifically, \textit{old} means that the node was in the network at least in the previous and in the current year. It does not directly correspond to constructed or demolished airports.} to the BAN also play an important r\^ole (Fig.~\ref{fig:03}). Nearly $18\%$ of the total routes in a year are new routes established betwen old airports (case aOO~\footnote{The code means: ``a'' or ``d'' for added (new) or deleted (removed) edges, and ``N'', ``R'' or ``O'' for new, removed or old airports respectively.}). The number of added routes between old and new airports is less than $8\%$ (aON). On the other hand, about $20\%$ of the routes from the previous year are deleted between old airports (dOO), while the number of routes added between old and removed airports is just about $6\%$ (dOR). The number of routes added between new airports (aNN) and deleted between removed airports (dRR) are irrelevant, approximately $0.2\%$ and $0.1\%$ (Fig.~\ref{fig:03}). This result reflects the natural tendency of airline companies to optimize ground staff and use the same facilities rather than investing heavily on routes to new airports.

Important to observe that only $100$ airports are identified in all years but $395$ different airports are counted during the period. Most of the airports appear and disappear during the years, but some of them stay in the network for a while and play an important r\^ole before being removed. The first case contains airports that temporary serve other nearby airports under makeover or even, unsuccessful tentatives of new routes. It also happens that small airports are included in the network for a short period of time. The airports in the second case are still important in terms of other services (e.g. air charter and private flights), however, due to several reasons, regular services are moved to specific cities in some regions.

This rewiring dynamics has important consequences in defining the paths between airports not directly connected. Since the average path length is larger than $1$ and the diameter can be at most $6$ (Table~\ref{tab:01}), one needs to go through intermediate airports to reach farther destinations and most important, sometimes the same aircraft and staff have to travel through many different airports during the day.

\begin{figure}[ht]
  \begin{center}
  \includegraphics[scale=0.45, angle=-90]{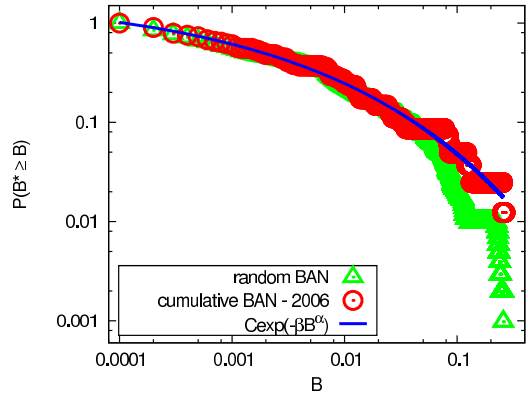} \\
  \end{center}
  \caption{(Color online) Cumulative betweeness distribution of the real and random BAN in $2006$. The data points are reasonable fitted by a stretched exponential with parameters $C=1.921$, $\beta=6.629$ and $\alpha=0.254$.}  \label{fig:05}
\end{figure}

The importance of the intermediate airports can be quantified by measuring the betweenness centrality (eqn.~\ref{eq:03}). The asymmetry in the relevance of some airports is observed on the (cumulative) betweenness distribution, which is relatively broad for all considered years. Figure~\ref{fig:05} shows that in $2006$ for instance, the cumulative distribution is reasonably approximated by the functional form $P(B^{*} \geqslant B) = C\exp(-\beta B^{\alpha})$ indicating that some airports are the most central ones. It suggests that most of the long range flights go through few intermediate airports which represent a socioeconomic relevance for a specific region and/or for the country itself. It is observed that the average betweenness has increased along the years (see SM -- Fig. 2). This result indicates that flights between non directly connected airports are going through specific intermediate airports, meaning that some long non-profitable routes are disappearing and being replaced by routes connecting more important airports. This situation has drawbacks because the network becomes more dependent on these central nodes. Any problem related to delays (e.g. bad weather or holidays) on key airports can propagate and disturb the functioning of the whole network. Moreover, since most of the connections go through them, they should be effective for control of, for instance, smuggling, drugs traffic or disease spread.

A path between two non-directly connected airports is not necessarily through the shortest path. Actually, many times alternative routes are available and normally, companies offer different travel options. Moreover, some companies (or even the passengers themselves) have the strategy of choosing cycles rather than going back and forth using the same sequence of airports. In a network, a cycle of size $m$ is defined as a closed path that goes from/to node $i$, passing through $m-1$ intermediate nodes, but not repeating a visited node. The BAN is observed to have more cycles than its random version with a relative peak in $1998$ for small cycles and in $2000$ for larger ones (see SM -- Table 1).

\subsection{Weighted BAN}

\begin{figure}[ht]
  \begin{center}
  \includegraphics[scale=0.45, angle=-90]{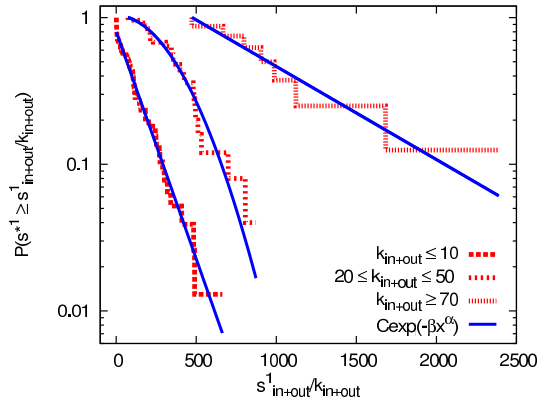} \\
  \end{center}
  \caption{(Color online) Cumulative distribution of the number of flights divided by the number of routes ($s_{in+out}^1/k_{in+out}$) for the BAN in $2006$. Data are separated on $3$ intervals, each curve is normalized independently. The best fits are given by the functional form $C$$\exp(-\beta x^{\alpha})$: $C=0.803$, $\beta=7.139\text{x}10^{-3}$ and $\alpha=1.0$ (for $k\leqslant10$); $C=1.033$, $\beta=5.409\text{x}10^{-6}$ and $\alpha=2.0$ (for $20\leqslant k\leqslant50$); and $C=2.004$, $\beta=1.461\text{x}10^{-3}$ and $\alpha=1.0$ (for $k\geqslant70$).}  \label{fig:06}
\end{figure}

The routes themselves do not explain all the dynamics of the flights, specially, because the routes have different importance in terms of connecting different airports. If we consider the strength $s^1_i$ of a node $i$ by summing the total number of connections (or flights -- layer $1$) during the year and divide it by the total number of routes (or (in+out)-degree), we can obtain the cumulative distributions in Figure~\ref{fig:06}. The resulting distributions corresponding to 3 different degree intervals are well described by the functional form $Cexp(-\beta x^{\alpha})$. The exponential parameter is larger for lower degrees and the $3$ intervals overlap on some regions. These results indicate a non-trivial pattern of physical/traffic constraints according to the airport size. Airports with many different routes (the so-called \textit{hubs}) have a lot of flights per route (at least $1.5$ flights per day) while most of the low-degree airports have much fewer flights per route. The former case correspond to the busiest airports in the BAN, while the last case includes small airports operating few regular flights (or routes not operated in a daily-basis but in some specific days during the week, which is due to several reasons, e.g. weekend flights from/to touristic places). On the other hand, airports lying in the intermediate range ($20\leqslant k\leqslant50$) appear to follow a different distribution with stronger decay ($\alpha=2$). It might include airports which started or ceased to operate during the year and mainly airports which decreased the number of flights on a specific route or simply eliminated a route.

\begin{figure}[ht]
  \begin{center}
  \includegraphics[scale=0.45, angle=-90]{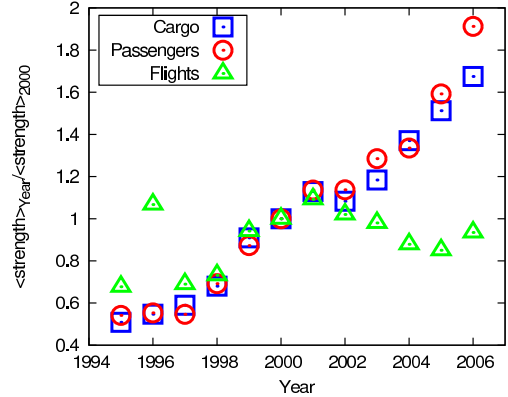} \\
  \end{center}
  \caption{(Color online) Average strength evolution on several layers in the BAN relative to year $2000$: layer $1$ representing the number of flights $\langle s^1\rangle$; layer $2$, the number of passengers $\langle s^2\rangle$; and layer $3$, the amount of cargo carried $\langle s^3\rangle$.}  \label{fig:07}
\end{figure}

Investigating the other two layers of the network, we observe that the traffic does not follow the shrink of the BAN. Indeed, the total traffic of passengers (layer $2$) and cargo (layer $3$) more than double in the period, from $18$ millions passengers in $1995$ to $43$ millions passengers in $2006$, and from $0.4$ megatons in $1995$ to $0.9$ megatons in $2006$. Figure~\ref{fig:07} shows that the average strength for each case (connections $\langle s_{in+out}^1\rangle$, passengers $\langle s_{in+out}^2\rangle$ and cargo $\langle s_{in+out}^3\rangle$) followed different dynamics, possibly converging to a higher occupancy level in the aircrafts (the capacity of the aircrafts increased as well). Important to observe that although this situation seems favorable to airline companies because of among other reasons, less maintenance costs (e.g. support staff or slots rent), actually, two large and old Brazilian airline companies ceased their activities\footnote{Indeed, one of the companies was sold and has undergone a restructuring process.} during this period and some of the routes were absorbed by other companies, two of them dominating most of the market nowadays.

\begin{figure}[ht]
  \begin{center}
  \includegraphics[scale=0.45, angle=-90]{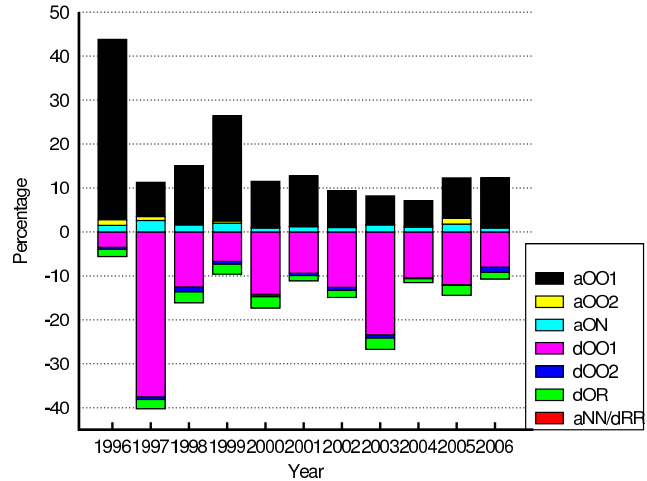} \\
  \end{center}
  \caption{(Color online) Percentage of added (a) (based in the current year) and deleted (d) (based in the previous year) connections between new (N), removed (R) and old (O) airports in the BAN. aOO1: added connections between old airports on existing routes; aOO2: added connections between old airports on new routes; aON: added connections between old and new airports; dOO1: deleted connections between old airports, but maintaining the route; dOO2: deleted connections between old airports, but removing the route; dOR: deleted connections between old and removed airports. aNN: added connections between new airports; and dRR: deleted connections between removed airports.}  \label{fig:08}
\end{figure}

The analysis of the added and deleted connections in the BAN provides useful insights about the dynamics of flights (Fig.~\ref{fig:08}). For the sake of organization, we define eight categories: added connections being introduced between old airports on existing routes (aOO1) or on new routes (aOO2); added connections introduced between an old and a new airport (aON) and finally, added connections between two new airports (aNN), which is very unlikely (less than $0.01\%$). In the case of removal of connections we have the situation were they are deleted from old airports, keeping the route (dOO1) or removing the route all together (dOO2); the case where connections are deleted between an old and a removed airport (dOR), or deleted between two removed airports (dRR), which is as unlikely as case $4$ (Fig.~\ref{fig:08}). Complementary to the routes analysis (Fig.~\ref{fig:03}), we observe that the BAN has large fluctuations in terms of the number of flights from year to year. Rather than trying to create new routes, companies follow the demand and invest on popular routes offering more flights. At the same time, connections tend to be removed before the route itself ceases to be operated. Moreover, it is reasonable to expect that when a new route is established, less frequent flights are offered to get a feedback before increasing the investment~\footnote{We might indicate possible external influences as the source of specific variations. For instance, the two most aggressive expansions in $1996$ and $1999$ are apparently followed by large removal of connections in the next year, meaning that aggressive investments do not necessarily increases profits. The federal government changed the policy and removed market restrictions at the end of $1997$~\cite{Febeliano:06}, which clearly affected the sector as we can see in the expansion of the number of connections in the two following years (Fig.~\ref{fig:08}). However, new regulation rules starting in $2003$~\cite{Febeliano:06} can be the responsible for the large removal of connections in this year. The substantial removal of connections in cases $6$ and $7$ (in special) in the year $2005$ can be related to problems with the previous cited airline companies which stopped operating.}.

These structural changes have strong impacts on single airports, with some of them becoming more central and others more peripheral with time. In terms of infra-structure, it means that the airports have different r\^oles in the network and consequently, policies related to new services and technologies have to be taken into account to avoid crashes of the whole system. The ten most important cities are selected and the respective centrality measurements are presented in Table~\ref{tab:03}. The number in parentheses shows the rank in the respective year and ``$\vartriangle$'' gives the variation based on the year $1995$.

We note that some cities are served by more than one airport (indicated by asterisks), for instance, the city of S\~ao Paulo is mainly served by two airports~\footnote{Officially, Governador Andr\'e Franco Montoro International Airport, which is in the neighbor city of Guarulhos and Congonhas International Airport, which is located inside the city of S\~ao Paulo but no longer used as an international airport.}. S\~ao Paulo is consolidated as the major hub in the BAN, although suffering as one of the cities which lost more routes, it has the largest increase in the number of connections during the period. It suggests that some profitable routes motivate the increase in the number of connections, while economical uninteresting routes were removed. Although the betweenness decreased $10\%$, it is still one of the most central cities in the network~\footnote{The evolution of S\~ao Paulo connections can be viewed in the support material.}. S\~ao Paulo airports are important for several reasons: S\~ao Paulo is the main connection between Brazil and countries from other continents, which means that many people necessarily go there before traveling abroad (indeed, S\~ao Paulo is ranked as the $16th$ most central airport in the world airport network in the year $2000$~\cite{Guimera:05}); the city itself is strategically located such that it connects the southern with the northern and the northeastern part of Brazil; and S\~ao Paulo is by far the most populous city in South America which demands a lot of traffic by all means of transportation.

\begin{table*}
\begin{center}
  \begin{tabular}{ | l || rl | rl | r | rl | rl | r | rl | rl | r | }
  \hline
               &\multicolumn{5}{c|}{$B$} &\multicolumn{5}{c|}{$k_{in+out}$} &\multicolumn{5}{c|}{$s_{in+out}^1$}\\ \hline
Airport        &\multicolumn{2}{c|}{1995}&\multicolumn{2}{c|}{2006}&\multicolumn{1}{c|}{$\vartriangle$} &\multicolumn{2}{c|}{1995}&\multicolumn{2}{c|}{2006}&\multicolumn{1}{c|}{$\vartriangle$} &\multicolumn{2}{c|}{1995}&\multicolumn{2}{c|}{2006}&\multicolumn{1}{c|}{$\vartriangle$}  \\ \hline
Manaus*        & $0.110$ & (3)  & $0.267$ & (1)  & $+143\%$ &  $87$ &(11) &  $85$ &  (5) & $-2\%$  &  $50,222$ &(11)  &  $40,081$ &(11) & $-20\%$ \\
S\~ao Paulo*   & $0.282$ & (1)  & $0.254$ & (2)  & $-10\%$  & $205$ & (1) & $145$ &  (1) & $-29\%$ & $282,578$ & (1)  & $346,495$ & (1) & $+23\%$ \\
Bel\'em*       & $0.070$ & (5)  & $0.133$ & (3)  & $+90\%$  &  $97$ & (8) &  $72$ &(7/8) & $-26\%$ &  $62,192$ & (6)  &  $47,995$ &(10) & $-23\%$ \\
Salvador       & $0.041$ &(12)  & $0.122$ & (4)  & $+198\%$ &  $95$ & (9) &  $95$ &  (3) &  $0\%$  &  $77,822$ & (5)  &  $93,757$ & (4) & $+20\%$ \\
Bras\'ilia     & $0.146$ & (3)  & $0.085$ & (5)  & $-42\%$  & $164$ & (2) & $103$ &  (2) & $-37\%$ & $100,449$ & (3)  & $115,550$ & (3) & $+15\%$ \\
Belo Horizonte*& $0.061$ & (7)  & $0.084$ & (6)  & $+38\%$  & $116$ & (4) &  $81$ &  (6) & $-30\%$ &  $61,925$ & (7)  &  $64,631$ & (7) & $+4\%$  \\
Rio de Janeiro*& $0.053$ & (8)  & $0.077$ & (7)  & $+45\%$  & $118$ & (3) &  $91$ &  (4) & $-23\%$ & $147,165$ & (2)  & $153,062$ & (2) & $+4\%$  \\
Curitiba*      & $0.090$ & (4)  & $0.035$ & (8)  & $-61\%$  & $111$ & (5) &  $65$ & (10) & $-41\%$ &  $57,674$ & (9)  &  $54,507$ & (9) & $-5\%$  \\
Fortaleza      & $0.069$ & (6)  & $0.032$ & (9)  & $-54\%$  &  $98$ & (7) &  $72$ &(7/8) & $-27\%$ &  $61,473$ & (8)  &  $65,115$ & (6) & $+6\%$  \\
Recife         & $0.012$ &(23)  & $0.018$ &(14)  & $+50\%$  &  $78$ &(14) &  $63$  &(11) & $-19\%$ &  $81,813$ & (4)  &  $73,386$ & (5) & $-10\%$ \\
\hline
\end{tabular}
\end{center}
\caption{The $10$ most important cities in the years $1995$ and $2006$. The measurements correspond to betweenness centrality ($B$), total number of routes ($k_{in+out}$) and total number of connections ($s_{in+out}^1$). In parentheses, the rank of the city in the respective year. ``$\vartriangle$'' gives the variation based on the year $1995$. Asterisks indicate cities served by $2$ or $3$ airports.} \label{tab:03}
\end{table*}

The interesting case of Manaus with a remarkable high betweenness is worth observing. This city is located in the middle of the Amazon rainforest (north of Brazil -- see Figure~\ref{fig:00} for the localization) and it has a central r\^ole to unite cities in that region with the other parts of the country. In general, most of the listed cities become more central with time (higher betweenness and more connections), indicating that though the network shrinks on the routes level, the increase in the amount of passengers and cargo is focused in specific profitable routes.

\section{Conclusions}

Transportation networks have a strategic r\^ole in society because people and goods are constantly moving between distances of different scales. These movements are fundamental for diversity in culture and behavior, and also diffusion of information. Meanwhile, trades between different regions keep the economy strong, but require a high level of efficient, profitable and trustful transportation networks. On the other hand, transportation networks also contribute for instance, to disease spread, drugs traffic and smuggling. The aviation sector has a special relevance in this context since it provides fast, long-range paths between different places.

Since the aviation sector is sensitive to economical fluctuations and geopolitical constraints, it is interesting to investigate how such network evolved during the recent years. By using the framework of complex networks theory, the structure and evolution of the BAN are investigated in several levels, from routes to passengers/cargo. The importance of the Brazilian airport network (BAN) lies in the fact that Brazil is an economically developing country with a large territory and sparse agglomerates of people.

It is observed that although some structural features are in accordance with other previous results from different airport networks, the BAN has a rewiring dynamics taking place at all levels. The results suggest that the companies have a tendency to invest in the most profitable routes rather than in new routes, consequently, increasing the number of connections on specific routes. The number of routes together with the number of airports decrease during the period, but the routes are constantly changing and not necessarily within the most connected airports. This dynamical evolution resulted on some airports becoming more central with time, while others become more peripheral. Specifically, the average betweenness of the network increased during the period and the degree distribution become more homogeneous. Broad degree and betweenness centralities (cumulative) distributions were observed in all years. Despite of the rewiring, the BAN showed a high number of cycles when compared to a random network, indicating that there are well defined internal structures on specific regions of the network. The structural changes in the ground levels had positive effects in the other levels, the number of passengers and the cargo carried through the network more than doubled in the considered period.

Complementary questions have been left open and might be of interest. For instance, we do not know whether this structural evolutionary path is a pure effect of the socioeconomic development of the country or an universal feature of this type of system. Indeed, it seems that other countries/regions also have a dynamic network evolution, for instance, the number of available seats decreased in the US network and increased in Asia and Europe networks in the last 10 years~\cite{OAG:08}. Moreover, specific system failures and attacks should be investigated more carefully since in fact, there is movement of carriers over these layers (e.g. congestion dynamics) and the respective time scale is different than the one investigated here (e.g. a delay occur within a period of one or few days).

\begin{acknowledgments}

The author acknowledges Luana de F. Nascimento, Matheus P. Viana, Sebastian Berhardsson, Petter Minnhagen, Luciano da F. Costa, Gonzalo Travieso and Petter Holme for important comments on earlier versions of the manuscript, and the Swedish Research Council (grant 621-2002-4135) for financial support.

\end{acknowledgments}


\begin{thebibliography}{10}

\bibitem{Colizza:06}
V.~Colizza, A.~Barrat, M.~Barth\'elemy, and A.~Vespignani.
\newblock The role of the airline transportation network in the prediction and
  predictability of global epidemics.
\newblock {\em Proceedings of the National Academy of Sciences}, 103(7):2015 --
  2020, 2004.

\bibitem{Boccaletti:review}
S.~Boccaletti, V.~Latora, Y.~Moreno, M.~Chaves, and D.-U. Hwang.
\newblock Complex networks: structure and dynamics.
\newblock {\em Physics Reports}, 424:175 -- 308, 2006.

\bibitem{Costa:review}
L.~da~F. Costa, F.~A. Rodrigues, G.~Travieso, and P.~R. Villas~Boas.
\newblock Characterization of complex networks: a survey of measurements.
\newblock {\em Advances in Physics}, 56:167 -- 242, 2007.

\bibitem{Costa2:review}
L.~da~F. Costa, O.~N. Oliveira~Jr, G.~Travieso, F.~A. Rodrigues, P.~R.
  Villas~Boas, L.~Antiqueira, M.~P. Viana, and L.~E. C.~da Rocha.
\newblock Analyzing and modeling real-world phenomena with complex networks: a
  survey of applications.
\newblock {\em e-print arXiv:physics/0711.3199}, 2007.

\bibitem{Barrat:04}
A.~Barrat, Barth\'elemy, R.~Pastor-Satorras, and A.~Vespignani.
\newblock The architecture of complex weighted networks.
\newblock {\em Proceedings of the National Academy of Sciences}, 101(11):3747
  -- 3752, 2004.

\bibitem{Barthelemy:05}
M.~Barth\'elemy, A.~Barrat, R.~Pastor-Satorras, and A.~Vespignani.
\newblock Characterization and modeling of weighted networks.
\newblock {\em Physica A}, 346:34 -- 43, 2005.

\bibitem{Guimera:05}
R.~Guimer\`a, S.~Mossa, A.~Turtschi, and L.~A.~N. Amaral.
\newblock The worldwide air transportation network: anomalous centrality,
  community structure and cities' global roles.
\newblock {\em Proceedings of the National Academy of Sciences}, 102(22):7794
  -- 7799, 2005.

\bibitem{Li:04}
W.~Li and X.~Cai.
\newblock Statistical analysis of airport network of {C}hina.
\newblock {\em Physical Review E}, 69:046106, 2004.

\bibitem{LiPing:03}
C.~Li-Ping, W.~Ru, S.~Hang, X.~Xin-Ping, Z.~Jin-Song, L.~Wei, and C.~Xu.
\newblock Structural properties of {US} flight network.
\newblock {\em Chinese Physics Letters}, 20(8):1393 -- 1396, 2003.

\bibitem{Bagler:04}
G.~Bagler.
\newblock Analysis of the airport network of {I}ndia as a complex weighted
  network.
\newblock {\em e-print arXiv:cond-mat/0409773}, 2004.

\bibitem{Chi:04}
L.~P. Chi and X.~Cai.
\newblock Structural changes caused by error and attack tolerance in {US}
  airport network.
\newblock {\em International Journal of Modern Physics B}, 18(17/19):2394 --
  2400, 2004.

\bibitem{Li:06}
W.~Li, Q.~A. Wang, L.~Nivanen, and A.~Le-M\'ehaut\'e.
\newblock How to fit the degree distribution of the air network?
\newblock {\em Physica A}, 368(1):262 -- 272, 2006.

\bibitem{Wang:05}
W.~Ru and C.~Xu.
\newblock Hierarchical structure, disassortativity and information measurements
  of the {US} flight network.
\newblock {\em Chinese Physics Letters}, 22(10):2715 -- 2718, 2005.

\bibitem{Han:07}
D-D. Han, J-H. Qian, and J-G. Liu.
\newblock Network topology of the {A}ustrian airline flights.
\newblock {\em e-print arXiv:physics/0703193}, 2007.

\bibitem{Guimera:04}
R.~Guimer\`a and L.~A.~N. Amaral.
\newblock Modeling the world-wide airport network.
\newblock {\em The European Physical Journal B}, 38:381 -- 385, 2004.

\bibitem{Hu:07}
Y.~Hu, D.~Zhu, and N.~Zhu.
\newblock A weighted network evolution model based on passenger behavior.
\newblock {\em e-print arXiv:physics/0709.4143}, 2007.

\bibitem{Barthelemy:06}
M.~Barth\'elemy and A.~Flammini.
\newblock Optimal traffic networks.
\newblock {\em Journal of Statistical Mechanics: Theory and Experiment},
  38:L07002, 2006.

\bibitem{Febeliano:06}
A.~Febeliano, C.~M\"uller, and A.~V. M.~de Oliveira.
\newblock The evolution of airline regulation in {B}rasil.
\newblock {\em Aerlines Magazine}, 33, 2006.

\bibitem{Freeman:79}
L.~C. Freeman.
\newblock Centrality in social networks conceptual clarification.
\newblock {\em Social Networks}, 1(3):215 -- 239, 1979.

\bibitem{Erez05a}
T.~Erez, M.~Hohnisch, and S.~Solomon.
\newblock {\em Economics: Complex Windows}, chapter Statistical Economics on
  Multi-Variable Layered Networks, page 201.
\newblock Springer, 2005.

\bibitem{anuar}
Anu\'ario estat\'istico do transporte a\'ereo.
\newblock Technical report, Ag\^encia Nacional de Avia\c{c}\~ao Civil,
  http://www.anac.gov.br (2008), 1995 -- 2006.

\bibitem{Wang:06}
B.~Wang, H.~Tang, G.~Chonghui, and Z.~Xiu.
\newblock Entropy optimization of scale-free networks' robustness to random
  failures.
\newblock {\em Physica A}, 363:591 -- 596, 2006.

\bibitem{OAG:08}
Oag revises its 4th quarter analysis of global airline activity based on
  updated published flight schedules.
\newblock Technical report, OAG, http://www.oag.com/oagcorporate/index.html,
  8th October 2008.

\end{thebibliography}
\end{document}